\documentstyle{aipproc}

\begin{document}
\title{Matter Fields in Curved Space-Time\footnote{Based on the talk given=
 by K.C.
Wali}}

\author{Nguyen Ai Viet\footnote{Permanent address: Department of High-Energy=
 Physics, Center
for Theoretical Physics, Institute of Physics, Hanoi, Vietnam} and
Kameshwar C. Wali}
\address{Physics Department, Syracuse University, Syracuse, NY 13244-1130}

\maketitle

\begin{abstract}
We study the geometry of a two-sheeted space-time within the
framework of non-commutative geometry. As a prelude to the
Standard Model in curved space-time, we present a model of a left-
and a right- chiral field living on the two sheeted-space time and
construct the action functionals that describe their interactions.
\end{abstract}

\noindent{\bf I. Introductory Remarks}

\medskip
%
%
%
%

It has been said by many, many times and in many words that our
current picture of space-time is unsatisfactory and inadequate for
the description of elementary particle interactions at all scales.
The reasons are many-fold. The most noted one is the fact that, in
spite of great deal of effort, the twin pillars of the twentieth
century physics, namely, the general theory of gravity (governing
the dynamics of classical space-time coupled to the dynamics of
matter fields in it) and quantum field theories (with the rules
quantization to be applied in principle to all degrees of freedom
including gravity) are found to be mutually incompatible. Attempts
over the past several decades at quantizing general relativity
have yet to meet with success. Superstring theory, as the
candidate for a consistent quantum theory of gravity along with a
unified description of all elementary particle interactions, has
remained so far only a promise. In spite of its recent new
insights through duality, of an underlying unity among a diversity
of string theories, described by a so called M-theory, is as yet
far from a convincing physical theory with predictable and
experimentally verifiable consequences.

The mathematical framework underlying both, the general theory of
relativity and quantum field theories, is based on a continuum
picture of space-time. A pseudo-Riemannian manifold endowed with a
metric structure based on a continuum picture underlies the
general theory of relativity. Likewise, quantum fields and their
interactions are local operators that are functions of continuous
commuting space-time coordinates. There are several reasons to
believe why such a continuum picture of space-time is untenable at
all distance scales. The problem of singularities in the curvature
tensor in general relativity and the ultra-violet divergences in
quantum field theories are too well known to merit discussion. To
these we might add two other problems that have received
considerable attention in recent years. One is the problem of
black hole entropy and the enumeration of the black hole degrees
of freedom and the other, the problem of localization of an event
consistent with quantum mechanics. To elaborate briefly on the
latter, we note that when we perform precise measurements of the
space-time localization of an event, up to uncertainties, $\Delta
X ^0, \Delta X^1, \Delta X^2, \Delta X^3,$ we must transfer energy
of the order $E \approx 1/\Delta X$. This energy creates a
gravitational field, and assuming spherical symmetry for
simplicity, the associated Schwarzschild radius $R \approx E
\approx 1/\Delta X$. Consequently signals originating inside $R$
cannot reach an outside observer. Such arguments point to the
existence of fundamental space-time uncertainty relations (STURS)
[1],

$$\Delta X^0 \Delta X \approx \ell_p^2\ \ ;\qquad\qquad \Delta X^1
\Delta X^2 \approx \ell_p^2$$

\noindent where $\ell_p$ is the Planck length.

Clearly such uncertainty relations are incompatible with classical
commuting coordinates of a Lorentzian or a Riemannian manifold.
They seem to imply [2] that the quantum theories of space, time
and matter are all interwoven. Hence, the ultimate goal of a
fundamental theory should be a generalized quantum theory, a
theory that does not, at the outset, begin with a continuum
space-time as an input, but a theory that gives rise to the
classical continuum of space-time in an appropriate limit.
However, at present, we have no real candidate for such a unified
quantum theory of space, time and matter, although there are
promising guideposts and indications of progress from several
different points of view.

One such promising approach has been provided by Alain Connes
based on, what is called, non-commutative geometry (NCG) [3,4].
Connes' ideas hinge upon the well-known theorem due to Gelfand
[3,5], which states that the classical topological space based on
a continuum can be completely recovered by the abelian algebra of
smooth functions defined on that space. As a natural
generalization, Connes considers non-commutative, but associative
and involutive algebras as the starting point for the description
of more general spaces with non-commuting coordinates, or spaces
with both continuous and discrete degrees of freedom. He
reconstructs the standard objects of the conventional differential
geometry in a purely algebraic way, setting up the basis for his
non-commutative geometry. It has given rise to the description of
the Standard Model with a geometrical interpretation of the Higgs
field on the same footing as the gauge fields with spontaneous
symmetry breaking following as a natural consequence.

In what follows, we present a brief account of work in progress,
which uses the basic ideas of Connes to construct Lagrangians and
action functionals of interacting matter fields in curved
space-time. This is an extension of our previous work [6] that
dealt with  the geometry of a two-sheeted space-time within the
framework of NCG and the consequent models for gravitational
interactions. In this report, we include matter fields. For
simplicity, we consider two chiral fields interacting with abelian
gauge fields. Their setting in curved space-time brings into play
the gravitational interactions and together they provide a rich
and complex model for further study.

In the next section, we discuss briefly the basic elements in
constructing the action funtionals in the usual Riemannian
geometry (R-geometry) to set the stage for their adoption in the
framework of non-commutative geometry. In Section III, we consider
a two-sheeted, non-commutative space-time with matter fields and
discuss the construction of action functionals involving
fermionic, gauge and gravitational interactions. The final section
is devoted to some concluding remarks.

\bigskip
\noindent{\bf II. Action functionals in R-Geometry; Basic ideas of
NCG}

\medskip

The starting point in the conventional R-geometry is the
construction of tangent and cotangent vector spaces $T_x(M)$ and
$T_x^*(M)$ at an arbitrary local point, $x$. They are spanned
respectively by the coordinate derivative operators
$\{\partial/\partial x^\mu\}$, and the coordinate differentials
$\{dx^\mu\}, \mu =3D 0,1,2,3,=85.,$ n-1, in an n-dimensional manifold,
so that an arbitrary tangent vector $X$ and an arbitrary cotangent
vector or one-form $\omega$ are linear combinations with real or
complex coefficients,

$$ X =3D X^\mu\partial_\mu\  ,\qquad\qquad  \omega =3D
dx^\mu\omega_\mu\ . \eqno (2.1)$$

These vector spaces are dual via the scalar product

$$dx^\mu(\partial_\mu) \equiv \langle dx^\mu\ ,\partial_\nu\rangle
=3D \delta_\nu^\mu,$$

\noindent and hence

$$\omega(X) \equiv \langle\omega, X\rangle =3D \langle
dx^\mu\omega_\mu, X^\nu\partial_\mu\rangle =3D \omega_\mu^* X^\nu.
\eqno (2.2)$$

\noindent Higher rank tensors $T$ of type $(p,q)$ and a
differential n-form, $\omega$, as totally antisymmetric tensor of
type $(n,o)$ are defined in the standard way,

$$T =3D T_{\nu_1...\nu_p}^{\mu_1...\mu_q}\ dx^{\nu_1}
\otimes\cdots\otimes dx^{\nu_p} \otimes \partial_{\mu_1}\cdots
\otimes\partial_{\mu_q},\eqno (2.3)$$

$$\omega =3D dx^{\mu_1} \wedge dx^{\mu_2}\cdots \wedge dx^{\mu_n}
\omega_{\mu_1\mu_2\cdots\mu_n}\ . \eqno (2.4)$$

\noindent With tensor and wedge products, we can define products
between tensors of different ranks and different n-forms, and
hence construct algebras over tensors and differential forms,
which we will denote as $\Lambda_x(M)$ and $\Omega_x(M)$
respectively:

$$\Lambda_x(M) =3D \oplus_{p,q} \Lambda_x^{(p,q)}(M)\ ;\ \
\Omega_x(M) =3D \oplus_n\Omega_x^n(M). \eqno (2.5)$$

\noindent The scalar product defined in (2.2) implies an
isomorphism between the tangent and cotangent vector spaces. This
isomorphism lends itself to the introduction of a metric leading
to the definition of the scalar products between two one- forms
(or equivalently between two tangent vectors),

$$g(dx^\mu, dx^\nu) =3D g^{\mu\nu}(x)\ , \qquad g^{\mu\nu}(x)
\epsilon C^\infty(M) \eqno (2.6)$$

$$g(\omega_1,\omega_2) =3D g(dx^\mu\omega_{1\mu},
dx^\nu\omega_{2\nu}) =3D (\omega_{1\mu})^*
g^{\mu\nu}\omega_{2\mu}.\eqno (2.7)$$

   While the local coordinate basis (2.1) is the customary basis
to describe curved space-time, it is advantageous for our purposes
to utilize locally flat, orthonormal tetrad system
$\{e_a,\theta^a\}$ defined by  the transformations, $e_a^\mu$,
called the vierbines,

$$e_a =3D e_a^\mu \partial_\mu\ , \qquad \theta^a =3D dx^\mu e_\mu^a\
,$$

\noindent with

$$e_a^\mu e_\mu^b =3D \delta_a^b \ , \qquad e_a^\mu e_\nu^a =3D
\delta_\nu^\mu. \eqno (2.8)$$

\noindent It then follows that the metric in tetrad system is
given by

$$g(\theta^a, \theta^b) =3D (e_\mu^a)^* g^{\mu\nu} e_\nu^b \equiv
\eta^{ab}, \eqno (2.9)$$

\noindent with the Lorentzian signature chosen to be $\eta$ =3D
diag(-1, 1, 1,=85,1).

By sesquilinearity the metric structure can be extended for the
vector spaces of higher forms. Of particular interest for us to
construct the action functionals is the two form $\Omega^2(M)$, in
which case, $\{\theta^a\Lambda\theta^b\}$ provide the basis
elements for the space of two-forms, and we define the scalar
product

$$\langle\theta^a\Lambda\theta^b,\ \theta^c\Lambda\theta^d\rangle
=3D \eta^{ad}\eta^{bc}-\eta^{ac}\eta^{bd}. \eqno (2.10)$$

\noindent The next important concept in R-Geometry is the
covariant derivative operator (or affine connection), =D1, that
assigns to each one-form $\omega\epsilon\Lambda^1(M)$, a (0, 2)
tensor field, $\nabla\omega$,

$$\nabla:\ \Lambda '(M) \to \Lambda '(M)\otimes \Lambda '(M)\ ,
\eqno(2.11)$$

\noindent where the image of $\nabla\omega$ on vector fields X, Y,
has the usual linearity properties. Skipping details that are too
well known in the literature, we note

$$\nabla\theta^a =3D \theta^b\otimes\omega_b^a =3D \theta^b
\otimes\theta^c\omega_{bc}^a\ , \eqno (2.12)$$

\noindent where $\omega_b^a$ are  the connection one-forms and
$\omega_{bc}^a$ the connection coefficients (Christoffel symbols
in coordinate frame). With this, the Cartan Structure Equations
defining torsion T, and curvature R are as follows:

$$\begin{array}{rl} T^a & =3D d\theta^a- \theta^b\omega_b^a\ , \\
\noalign{\vskip 4pt}%
R_b^a & =3D d\omega_b^a + \omega_c^a \wedge \omega_b^c\ .\end{array}
\eqno (2.13)$$

\noindent As well known, if torsion is assumed to vanish, the
connection coefficients can be expressed in terms of the vierbeins
and their derivatives( that is, in terms of metric components and
their derivatives). Starting with appropriate one-forms, we can
then calculate the required curvature tensors using the second of
the Cartan structure equations. For the abelian gauge sector, for
instance, the one-form

$$A =3D dx^\mu A_\mu \eqno (2.14)$$

\noindent leads to the curvature

$$F =3D dA + A\wedge A\ , \eqno (2.15)$$

\noindent and the Lagrangian and the action

$${\cal L}_A =3D \langle F,F\rangle\ ; \ S_A =3D \int d^4 x \sqrt{-g}
\langle F,F\rangle .\eqno (2.16)$$

\noindent Likewise, in the gravitational sector, the connection
one-forms $\omega_b^a$ are the Riemannian connections in (2.13),
leading to the Riemannian curvature two-forms

$$R_b^a =3D \frac{1}{2} R_{bcd}^a \theta^e \wedge \theta^d\ , \eqno
(2.17)$$

\noindent and the Einstein-Hilbert Lagrangian and the action

$${\cal L}_{EH} =3D \langle R_{ab},
\theta\stackrel{a1}{\wedge}\theta^b\rangle\ , \ S_{EH} =3D \int d^4x
\frac{\sqrt{-g}}{16\pi G} R \eqno (2.18)$$

\noindent where

$$R =3D \frac{1}{2} (\eta^{bc} R_{bcd}^d-\eta^{bd}R_{bcd}^c)\ .$$

\noindent Next to obtain an action functional for a fermionic
field in curved space-time, consider a  Dirac spinor, $\Psi(x)$,
at an arbitrary point $x\epsilon M$, and a homomorphism $\gamma$
on the Clifford algebra $C(1,3)$ with the multiplication defined
on the basis one-forms

$$\theta^a\theta^b + \theta^b\theta^a =3D 2\eta^{ab}\ .$$

\noindent This provides the usual $\gamma$-matrices in flat
space-time with

$$\gamma(\theta^a) =3D \gamma^a \ , \qquad \{\gamma^a,\gamma^b\} =3D 2
\eta^{ab} 1\ . \eqno (2.19)$$

\noindent The $\gamma$-matrices in curved space-time are then
given by

$$\gamma^\mu(x) =3D e_a^\mu(x)\gamma^a\ , \qquad
\{\gamma^\mu,\gamma^\nu\} =3D 2 g^{\mu\nu} 1. \eqno (2.20)$$

\noindent Taking into account the correct transformation
properties, with the derivative operator, $\partial_a$, in flat
space-time replaced by, $\nabla_a$, where

$$\nabla_a =3D e_a^\mu(\partial_\mu + ie A_\mu + \frac{1}{2i}
e_\mu^a\omega_{abc}\sigma^{bc}), \eqno (2.21)$$

\noindent we have from the Dirac Lagrangian in flat space-time,

$${\cal L}_D =3D i \bar{\Psi}\gamma^a\partial_a\Psi\ ,$$

\noindent transformed into the action in curved space-time that
includes gauge and gravitational interactions,

$${\cal L}_D \mapsto {\cal L}_\Psi =3D i \bar{\Psi} e_a^\mu
(\partial_\mu + ie A_\mu + \frac{1}{2i} e_\mu^a
\omega_{abc}\sigma^{bc})\ \Psi. \eqno (2.22)$$

\noindent To summarize, the basic ingredients to construct matter
fields and their interactions including gravity  in R-geometry
are, the algebra of differential forms, $\Lambda^*(M)$, the affine
connection, $\nabla$, curvature( or field strengths) and an inner
product on tangent space, whose image is a Lorentz-invariant
scalar. These geometrical constructs depend strictly upon the
local properties defined in the neighborhood of a point in the
manifold. To generalize these notions to non-commutative spaces,
one needs to depart from this localization. In Connes version of
NCG, this is accomplished by first reformulating the above
concepts in algebraic terms and then an operator-theoretic
representation of the algebra on a Hilbert space. Thus, it
involves three basic elements $({\cal A,H,D})$,which are
collectively called a spectral triple.  In what follows, we shall
describe the main ideas to be applied to a two-sheeted space-time
in the following section. For a more extensive treatment of these
topics, see [2,6].

A). The Manifold $M \to {\cal A}$, an associative, involutive,
commutative or noncommutative algebra with a unit element, whose
elements are to be  represented as operators on a Hilbert space
${\cal H}$. A universal differential algebra, $\Omega^*{\cal A}$,
is generated by all $a \epsilon {\cal A}$ , and a symbol $\delta$,
such that

$$\delta(1) =3D 0\ ; \delta(ab) =3D (\delta a)b + a(\delta b), \forall
a,b\epsilon {\cal A}\ . \eqno (2.23)$$

\noindent By definition, the algebra of zero forms is
$\Omega^\circ{\cal A} \equiv {\cal A}$; $\delta a$ belongs to the
space of universal one-forms $\Omega^\prime {\cal A}$, whose
general element is a linear combination

$$ \omega =3D \Sigma_i \delta a_i b_i\ , \qquad a_i,b_i \epsilon
{\cal A}\ . \eqno (2.24)$$

\noindent In NCG, the natural generalization of a vector space of
one-form is an ${\cal A}$-module, where the  linear combinations
with real or complex coefficients are replaced by linear
combinations with coefficients belonging to the algebra. Likewise
scalar multiplication is defined with elements of the algebra from
the right (left) for right-(left-) ${\cal A}$-modules. Using the
Leibnitzian property of $\delta$, and repeated multiplication of
one-forms, we can construct higher p-forms on the algebra, and can
multiply two such arbitrary forms, generating a graded universal
algebra, $\Omega^*{\cal A} \equiv \oplus \Omega^p{\cal A}$,
corresponding to $\Omega_{x}(M)$ in (2.5 ). Further, by defining
$\delta$ as a linear operator,

$$\Omega^p \to \Omega^{p + 1}\ ,$$

$$\delta(\delta a_1\cdots\delta a_p b) =3D \delta a_1\cdots \delta
a_p \delta b\ ,$$

\noindent we can transform the graded algebra of forms,
$\Omega^*{\cal A}$, into a differential algebra.

\noindent Next, we note that in the standard treatment, the
physical matter fields are described as sections of vector and
covector bundles. The space of such sections on a continuum
manifold is always a finite, projective module over the
commutative algebra of smooth functions. We assume that in their
generalization in NCG, they are replaced by finite, projective
${\cal A}$-modules.

\bigskip

B). A self-adjoint operator, called the Dirac operator, ${\cal
D}$, on the Hilbert space ${\cal H}$, such that the commutator
[${\cal D}, a$] is a bounded operator, $\forall a \epsilon {\cal
A}$.  The objects of the universal differential algebra are then
represented as operators on by the following graded homomorphism,

$$\Pi : \Omega^*{\cal A} \to {\cal L}({\cal H})\ ,$$

$$\Pi_p(\delta a_1\cdots\delta a_pb) =3D
\begin{array}{ccc}_p\\ \Pi\\ \noalign{\vskip -6pt}%
 _{i=3D1}\end{array} [{\cal D}, \Pi_0(a_i)]\Pi_0(b) \eqno
(2.25)$$

\noindent where ${\cal L}({\cal H})$ denotes the space of bounded
operators on ${\cal H}$ and $\Pi_0$, the representation of ${\cal
A}$ on ${\cal H}$.

In the case of a 4-dimensional, pseudo-Riemannian, spin manifold,
M, using the canonicle  spectral triple, where ${\cal A} =3D
C^\infty(M,R), {\cal H} =3D L^2(M,S)$ the space of square integrable
sections of the spinor bundle, and $D =3D \gamma^\mu\nabla_\mu$, we
can reproduce the action functionals (2.16, 2.18, and 2.22),
derived using standard techniques [7].

\bigskip
\noindent{\bf III. Two-sheeted Space-time}

\medskip
A two sheeted space-time may be interpreted as a Kaluza-Klein
theory with an internal space of two discrete points in the fifth
dimension. As stated earlier, apart from providing a non-trivial
extension of R-geometry and an example of a noncommutative space,
it is also physically motivated from the fact that, due to parity
violation, there is an intrinsic difference between left- and
right-chiral fields. We might imagine that they live on two
separate copies of space-time. For simplicity, we consider here
two chiral spin-1/2 fields, each coupled to two an abelian gauge
field in the presence of gravity. Extension to the full Standard
Model will be dealt with elsewhere. In what follows, we shall use
M, N, .., and A, B,=85, to mark the indices in curvilinear frame and
the locally flat tetrad system respectively, with $M =3D \mu, 5
(\mu$ =3D 0,1,2,3) and $A =3D a,$\.{5}$(a =3D $0,1,2,3).

\noindent As discussed in the previous section, the basic
ingredient in NCG is the spectral triple $({\cal A,H,D})$. The
spectral triple for our model is as follows:

$${\rm Algebra} : \qquad {\cal A} =3D C^\infty(M)\otimes(R\oplus R)
=3D C^\infty(M,R) \oplus C^\infty(M,R)$$

\noindent with the elements of the algebra  given by

$$F(x) =3D f_+(x)\left(\begin{array}{ccc} 1 & 0\\0 &
1\end{array}\right) + f_-(x)\left(\begin{array}{ccc} 1 & 0 \\ 0 &
-1\end{array}\right) =3D \left(\begin{array}{ccc}f_1(x) & 0\\ 0 &
f_2(x)\end{array}\right)\ . \eqno (3.1)$$

$${\rm Hilbert\ Space}: \qquad {\cal H} \stackrel{\cdot}{=3D}
L^2(M,S)\oplus L^2(M,S)\qquad\qquad\qquad$$

\noindent consisting of left and right, square-integrable sections
of a spinor bundle. Elements of the algebra, ${\cal A}$, act as
operators,

$$F(x) \Psi(x) =3D \left(\begin{array}{ccc} f_1(x) & 0\\ 0 &
f_2(x)\end{array}\right)\left(\begin{array}{ccc}\psi_L(x)\\
\psi_R(x)\end{array} \right)\ . \eqno (3.2)$$

$${\rm Dirac\ Operator}: \qquad {\cal D} =3D \Gamma^M D_M =3D
\Gamma^\mu D_\mu + \Gamma^5 D_5\ ,\qquad\qquad\qquad$$

\noindent where the derivative operators $D_\mu ,D_5$ are given by

$$D_\mu =3D \left(\begin{array}{ccc}\nabla_\mu & 0\\ 0 &
\nabla_\mu\end{array}\right)\ \ , \ \ D_5 =3D
\left(\begin{array}{ccc}m & 0\\ 0 & m\end{array}\right) \eqno
(3.3)$$

\noindent The parameter m in (3.3) has the dimensions of mass to
conform to the dimensions in $D_\mu$.

\noindent In complete analogy with the spinor formulation of
R-geometry, we introduce two sets of gamma matrices, $\Gamma^A$
(locally flat space-time), and $\Gamma^M$ (curved space-time),

$$\Gamma^a =3D \left(\begin{array}{ccc} \gamma^a & 0\\ 0 &
\gamma^a\end{array}\right)\ , \ \ \Gamma^{\dot{5}} =3D
\left(\begin{array}{cc} 0 & \gamma^5\\ \gamma^5 &
0\end{array}\right)$$

\noindent and

$$\Gamma^M =3D \Gamma^A E_A^M\ , \eqno (3.4)$$

\noindent where $E_A^M$ are the generalized vierbeins that define
the metric. Without any loss of generality, we can choose them to
be

$$E_\mu^a(x) \dot{=3D} \left(\begin{array}{ccc} e_{1\mu}^a(x) & 0 \\
0 & e_{2\mu}^a(x)\end{array}\right)\ , \ \ E_5^a(x) \dot{=3D} 0$$

$$E_5^{\dot{5}}(x) \dot{=3D} \Phi(x) =3D
\left(\begin{array}{ccc}\phi_1(x) & 0 \\ 0 &
\phi_2(x)\end{array}\right)\ , \ \
\begin{array}{ll}E_\mu^{\dot{5}}(x) & =3D \left(\begin{array}{ccc}
a_{1\mu}(x) & 0 \\ 0 & a_{2\mu}(x)\end{array}\right) \Phi(x),\\
 & \dot{=3D}  A_\mu\Phi \ . \end{array} \eqno (3.5)$$

\noindent The orthogonality relations

$$E_M^A E_A^N =3D \delta_M^N\ , \ \ E_M^A E_B^M =3D \delta_B^A$$

\noindent then determine

$$\begin{array}{lll}E_a^\mu(x) =3D (E_\mu^a(x))^{-1} \ & , E_a^5(X)
=3D -E_A^\mu A_\mu\\
\noalign{\vskip 4pt}%
 E_{\dot{5}}^\mu (x) =3D 0 \ & ,
E_{\dot{5}}^5(x) =3D \Phi^{-1}(x).\end{array}\eqno (3.6)$$

The Dirac operator takes the form

$$D =3D \Gamma^M D_M =3D \left(\begin{array}{ccc}
e_{1a}^\mu\gamma^a\nabla^\mu & \gamma^5m\\ \gamma^5m &
e_{2a}^\mu\gamma^a\nabla_\mu\end{array}\right)\ . \eqno (3.7)$$

\noindent The generalized metric has the form

$$\begin{array}{lll} G^{MN}& =3D & E_A^M \eta^{AB} E_B^N =3D
\left(\begin{array}{ccc}G^{\mu\nu} & G^{\mu 5}\\ G^{5\mu} &
G^{55}\end{array}\right) ,\\ G_{MN} & =3D & E_M^A\eta_{AB} E_N^B =3D
\left(\begin{array}{ccc} G_{\mu\nu} & G_{\mu 5}\\ G_{5\mu} &
G_{55}\end{array}\right)\ ,\end{array}$$

\noindent where

$$G^{\mu\nu}(x) =3D \left(\begin{array}{ccc} g_1^{\mu\nu} & 0 \\ 0 &
g_2^{\mu\nu}(x)\end{array}\right)\ , \ \  G^{\mu 5} =3D G^{5\mu} =3D
-A^\mu , G^{55} =3D A^2 + \Phi^{-2}$$

\noindent and

$$G_{\mu\nu}(x) =3D  \left(\begin{array}{ccc} g_{1\mu\nu}(x) & 0
\\ 0 & g_{2\mu\nu}(x)\end{array}\right) +
A_\mu(x)A_\nu(x)\Phi^2(x), G_{\mu 5} =3D G_{5\mu} =3D
A_\mu(x)\Phi^2(x), $$

\vskip -15pt

$$G_{55}(x) =3D \Phi^2(x).\hspace {4in} \eqno (3.8)$$

\noindent Thus we find that, to describe the two-sheeted
space-time, we need in general, two vierbeins (i.e., two tensor
fields), two vector fields and two scalar fields. We shall call
them collectively as metric fields.

\noindent The generalized one-forms are give by

$$U =3D \Gamma^A U_A =3D \Gamma^M E_M^A U_A =3D \Gamma^M U_M\eqno
(3.9)$$

\noindent with $U_M$ being the elements of the algebra. Further,
as a direct generalization of the covariant derivative operator in
the standard R-geometry, we define

$$\nabla\Gamma^A =3D \Gamma^B \otimes \Omega_B^A\ , $$

\noindent where $\Omega_B^A$ are the generalized connection
one-forms corresponding to $\omega_b^a$ in (2.12).

The Cartan structure equations generalize easily to take the form

$$\begin{array}{ccc} T^A & =3D & D\Gamma^A +
\Gamma^B\wedge\Omega_B^A\ ,\\
\noalign{\vskip 4pt}%
 R_B^A & =3D & D\Omega_B^A +
\Omega_C^A \wedge\Omega_B^C\end{array}. \eqno (3.10)$$

 The connection one-forms are arbitrary
functions to be specified or to be determined by additional
conditions. As noted previously, in R-geometry, metric
compatibility and the condition that torsion vanish lead to the
determination of these forms and the associated scalars in terms
of the metric. We have shown in our previous work [6], a direct
generalization of these conditions in the present case leads to
constraints on the metric fields in the form

$$ e^a_{1 \mu} (x) =3D \beta (x) e^a_{2 \mu} (x)\ ,\ a_{1 \mu} (x) =3D
a_{2 \mu} (x) \alpha (x)\ ,\ \phi_1 (x) =3D {\phi_2 (x) \over \alpha
(x)} \eqno(3.11) $$

\noindent where $\beta (x)$ and $\alpha (x)$ are arbitrary
functions. With $\alpha =3D \beta =3D 1$, we were able to reproduce
the exact zero mode of the Kaluza-Klein theory. Seeking a more
general formulation that contains all the metric component fields
independently and a Lagrangian and an action that incorporates the
full set of tensor, vector and scalar fields, Ai Viet has found a
set of minimal constraints [8],

\medskip

\noindent a) Metric compatibility

$$\nabla G =3D 0 \quad \Rightarrow \quad \Omega^+_{AB} =3D -
\Omega_{BA}$$

\medskip

\noindent b) Torsion components involving the 4-dimensional
space-time components vanish

$$T^a =3D 0$$

\noindent as the natural generalization of the R-geometry

\medskip

\noindent c) The condition

$$\Omega_{AB \dot 5} =3D 0$$

     With the above conditions, he has shown that the rest of the
connection one-forms and the non-vanishing components of torsion
are determined in terms of metric fields (analogous to the
determination of the connection coefficients in terms of the
metric in R-geometry). This defines our gravity sector with six
independent metric fields:

$$E^\mu_{\pm a} (x) =3D {e^\mu_{1a} (x) \pm e^\mu_{2a} (x) \over 2}\
,\ A_{\pm \mu} (x) =3D {a_{1 \mu} (x) \pm a_{2 \mu} (x) \over 2}\
,$$

$$\Phi_{\pm} (x) =3D {\phi_1(x) \pm \phi_2 (x) \over 2}
\eqno(3.12)$$

\noindent While the full exploration of the ensuing general theory
of gravity is in progress, we would like to point out one general
feature. It is that one set of the component tensor, vector and
scalar fields is massless where as the other is massive. There are
essentially two parameters, the Newton's constant G and the
parameter m of the dimensions of mass. Next,taking this as the
background curved space-time, we proceed to construct the
Lagrangians for the gauge and fermionic sectors. The resulting
expressions are too long and involved to give their full forms
here. We present here only their qualitative features, leaving the
full details in papers to be published [7,8].
\bigskip

\noindent {\bf Gauge Sector} \bigskip

With the one-form

$$B =3D \Gamma^M B_M =3D \Gamma^A B_A \ ,\ B_A =3D E^M_A B_M\ ,$$

\noindent where

$$B_\mu (x) =3D \pmatrix{b_{1 \mu}(x) &0\cr
\noalign{\vskip 4pt}%
 0 &b_{2 \mu} (x)\cr}\ ,\
B_5 (x) =3D \pmatrix{h_1 (x)&0\cr
 \noalign{\vskip 4pt}%
 0 &h_2 (x)\cr} =3D H (x)
\eqno(3.13)$$

\noindent we have two abelian gauge fields and two Higgs fields.
The curvature given by

$$G(x) =3D DB (x) + B(x) \wedge B(x) =3D \theta^A \wedge \theta^B
G_{AB} \eqno(3.14)$$

\noindent  and hence the Lagrangian

$$<G,G>\  =3D G_{AB} G_{CD} \left( \eta^{AD} \eta^{BC} - \eta^{AC}
\eta^{BD} \right) \ . \eqno(3.15)$$

\noindent Written in terms of the linear combinations $B_{\pm \mu}
=3D {b_{1 \mu} \pm b_{2 \mu} \over 2}$, one finds that $b_{+ \mu}$
is massless, whereas $b_{- \mu}$ is massive. Proper kinetic terms
for the gauge and Higgs fields and Lorentz covariant interactions
ensure that we have a physically meaningful Lagrangian. In
addition to the expected gauge and Higgs fields interactions, it
has the new feature involving the interactions with vector and
scalar components, $a_\mu,\ \phi$, that are parts of the metric.

 We have a Higgs potential in the form

$$V(x) =3D \left( \Phi \tilde \Phi \right)^{-2} \left( \tilde H H +
m ( H + \tilde H) \right)^2 \eqno(3.16) $$

\noindent which with $h_1 =3D \overline h_2$ reduces to the required
standard form for spontaneous symmetry breaking, namely,

$${1 \over 2}\ V(x) =3D \left( \phi_1 \phi_2 \right)^{-2} \left(
\eta (x) \overline \eta (x) - m^2 \right)^2 \ ,$$

\noindent where

$$\eta =3D \eta_1 + i \eta_2 \ ,\  \eta_1 =3D {h_1 + h_2 \over 2} - m
\  , \ {\eta_1 =3D {h_1 + h_2 \over 2} - m}\ ,\ \eta_2 =3D {h_1 - h_2
\over 2i} \ . \eqno(3.17)$$

\bigskip

\noindent {\bf Fermionic Sector}
\bigskip

Beginning with the generalized Lagrangian,

$${\cal L}_F =3D i \overline \Psi \Gamma^A \left( E^M_A (D_M + B_M)
- {1 \over 4}\ \Gamma^B \Gamma^C \Omega_{BCA}\right) \Psi\ ,
\eqno(3.18)$$

\noindent we find in it the expected part,

\begin{eqnarray*}
& & i \overline \Psi \left\{ \gamma^a \left( e^\mu_{+a} \left(
\partial_\mu + b_{+ \mu}\right)+ e^\mu_{-a} b_{- \mu} \right) - {1 \over
8}\ \gamma^b \gamma^c \left( \omega_{1bca} + \omega_{2 bca}
\right)\right.\\
\noalign{\vskip 4pt}%
 & & \left. \quad + \gamma^5 \gamma^a  \left( e^\mu_{-a} \left(
\partial_\mu + b_{+ \mu}\right)+ e^\mu_+ b_{- \mu} \right) - {1 \over
8}\ \gamma^b \gamma^c \left( \omega_{1bca} - \omega_{2 bca}
\right)\right\} \Psi\ ,
\end{eqnarray*}

\noindent which is the Dirac Lagrangian in the curved space-time
of our model. It contains host of other terms arising from the
discrete part involving the Higgs fields,  $h_1,\ h_2$,  and  the
vector and scalar components $a_1,\ a_2$ and $\phi_1,\ \phi_2$
respectively. Assuming $e^\mu_{+a}$ defines the physical metric,
it is interesting to note that the massless $b_{+ \mu}$ has only
the vector interaction, whereas the massive $b_{- \mu}$ has both
vector and axial vector interactions. Thus, in our simplified
model of two chiral fields, $b_{+ \mu}$ represents the parity
conserving photon and $b_{- \mu}$
         represents the massive parity violating Z-vector boson. We
also note the following terms of special interest

\begin{eqnarray*}
& & i \overline \Psi \left[ {1 + \gamma_5 \over 2}\ (m + h_2)
\phi^{-1}_2 - {1 - \gamma_5 \over 2}\ (m+h_1) \phi^{-1}_1 \right]
\Psi\\
\noalign{\vskip 4pt}%
 & &  - i \overline \Psi \left[ {1 + \gamma_5 \over 2}\ (m +
h_1) \gamma^a a_{1a} + {1 - \gamma_5 \over 2}\ (m+h_2) \gamma^a
a_{2a}  \right] \Psi
\end{eqnarray*}

\noindent that involve interactions with the vector and scalar
components of the gravity sector. They violate parity, and
consequently raise the intriguing possibility of gravitation as
the origin or part of the origin of parity violation.

\bigskip

\noindent {\bf IV. Concluding remarks}

\bigskip

In this brief account of work in progress, we have presented an
extension of the usual Riemannian geometry to a two-sheeted
space-time, that is, a continuum space-time with a discrete
two-point internal space. As a simple but nontrivial example of a
noncommutative space, it has provided an extremely valuable model
to study non-commutative geometric approach of Connes as well as
the interplay of gravitational and other elementary particle
interactions. Additional strong physical motivation for our model
arises from the fact that  parity violation implies different
physical quantum numbers for left and right chiral fields in the
Standard Model, so that we can imagine the two chiral fields live
on two copies of space-time.

We have noted in our previous work, that this discretized
Kaluza-Klein theory with a finite field content generates a finite
mass spectrum. The appearance of new vector and scalar interaction
terms with correlated strengths and nonlinear in nature lends
itself to the speculation of possible softening of divergences in
the gravitational sector and their effect on renormalization. As
an extended theory of gravity, it has obviously cosmological
implications. When matter fields are introduced, we find that the
NCG framework allows spontaneous symmetry breaking and leads to a
rich and complex variety of interactions that certainly merit a
great deal of further study. Of particular importance is the
appearance of parity violating interactions due to gravitational
vector and scalar fields.

\bigskip

\noindent {\bf Acknowledgments}

\bigskip

This work was supported in part by the U.S. Department of Energy
under contract No. DE-FG02-85ER40231.

\bigskip

\noindent{\bf References}

\begin{enumerate}

\item S. Doplicher, K. Fredenhagen, and J. E. Roberts, Physics
Letters B 331 (1994) 39-44.

\item J. Frohlich, in PASCOS 94-Proc.IV Int. Symp. On Particles,
Strings and Cosmology (Syracuse, 1994),ed. K.C. Wali ( World
Scientific, Singapore, 1995), pp. 443

\item A. Connes, Non-Commutative Geometry (Academic Press, 1994)

\item A. Connes, " Essay on Physics and Non-Commutative Geometry,"
The Interface of Mathematics and Particle Physics, D.G. Quillen,
G.B. Segal, S.T. Tsou editors (Clarendon Press, Oxford, 1990)

\item G.F. Simmons, Introduction to Topology and Modern Analysis (
Krieger, 1983)

\item N.A. Viet and K.C.Wali, Intl.J.Modern Physics. A, 11 (1996),
pp. 533-551 and ibid. A, 11 (1996), pp. 2403-2418

\item Construction of Action Functionals in Non-Commutative
Geometry, J. A. Javor, N.A. Viet, and K.C. Wali ( SU 4240-692)

\item A New Minimal Set of Constraints on the Connections and
Torsion in Non-Commutative Geometry, N. A. Viet ( SU-4240-566)

\end{enumerate}

\end{document}